\documentclass[aps,pre,showpacs,longbibliography]{revtex4-1}
\usepackage{color}
\usepackage{amssymb}
\usepackage{amsmath}
\usepackage{bm}
\usepackage{graphicx}

\begin{document}
\title{Scaling of Rayleigh-Taylor mixing in porous media}

\author{G. Boffetta}

\author{M. Borgnino}

\author{S. Musacchio}

\affiliation{Dipartimento di Fisica and INFN, Universit\`a di Torino, 
via P. Giuria 1, 10125 Torino, Italy}

\begin{abstract}
Pushing two fluids with different density one against the other 
causes the development of the Rayleigh-Taylor instability at their
interface, which further evolves in a complex mixing layer.  
In porous media, this process is influenced by the viscous resistance
experienced while flowing through the pores, which is described by the
Darcy's law.
Here, we investigate the mixing properties of the Darcy-Rayleigh-Taylor system
in the limit of large P\'eclet number by means of direct numerical 
simulations in three and two dimensions.
In the mixing zone, the balance between gravity and viscous forces
results in a non-self-similar growth of elongated plumes,
whose length increases linearly in time while their width follows
a diffusive growth.
The mass-transfer Nusselt number is found to increase 
linearly with the Darcy-Rayleigh number supporting a
universal scaling in porous convection at high Ra numbers.
Finally, we find that the mixing process displays important 
quantitative differences between two and three dimensions.
\end{abstract}


\maketitle 

Rayleigh-Taylor (RT) is a well know instability at the interface of two 
fluids with different density in presence of a relative acceleration.
The instability develops in a
nonlinear phase with a mixing layer which grows in time. 
Usually, RT mixing is studied for bulk fluids at low viscosity where
the turbulent mixing layer grows as $t^2$ and produces
velocity and temperature fluctuations which
 follow the Kolmogorov-Obukhov phenomenology in three
dimensions (3D) \cite{boffetta2009kolmogorov} and the Bolgiano-Obukhov
phenomenology in two dimensions (2D) 
\cite{chertkov2003phenomenology,celani2006rayleigh,boffetta2012bolgiano}.
RT turbulence has been extensively studied within the general framework
of turbulent convection, and is today one of the prototype of buoyancy
driven flow with a wide range of applications
\cite{sharp1984overview,boffetta2017incompressible,zhou2017rayleigh}.

RT instability has also been studied in porous media, starting from 
the classical works on linear stability analysis \cite{saffman1958penetration}
and the comparison with experiments in quasi-2D Hele-Shaw cells
\cite{wooding1969growth}.
Recently, RT mixing in porous media has received a renewed attention in 
view of its applications, in particular in $CO_2$ sequestration in 
saline aquifers. 
Dissolution of carbon dioxide in the aqueous phase increases its density 
and induces a buoyancy-driven instability which accelerate the process of 
$CO_2$ sequestration into the bulk of the aquifer 
\cite{huppert2014fluid,ennis-king2005role,emami-meybodi2015convective}.
Motivated by this application, several experimental and numerical 
studies of RT mixing have been reported,
mostly in two dimensions
\cite{gopalakrishnan2017relative,depaoli2019rayleigh}
while few experiments in three dimensions have been performed
\cite{nakanishi2016experimental}.

In spite of their importance for applications, buoyancy driven flows in
porous media are much less known than their turbulent analogue in 
bulk flow. RT mixing is not an exception and indeed its basic statistical
properties are still not fully understood. 
In this Rapid Communication we study porous RT mixing by means of extensive
three-dimensional direct numerical simulations (DNS) of Darcy's equation 
at different resolutions and diffusivities
in order to achieve a possible asymptotic regime.
By comparing the results of 3D and 2D numerical simulations
at the same P\'eclet number we find similar qualitative behavior 
but significant quantitative differences.
We find that in 3D
the extension of the mixing layer grows linearly in time,
in agreement with dimensional predictions
and experimental findings \cite{wooding1969growth}
and in contrast with what observed recently in two dimensions
where anomalous scaling has been reported
\cite{depaoli2019rayleigh}.
The horizontal characteristic scale of plumes is found to follow a 
diffusive law and therefore the mixing layer does not evolve 
in a self-similar way. 
We also find that the dimensionless mass flux,
quantified by the Nusselt number,
grows proportionally to the Darcy-Rayleigh number of the flow 
both in two and three dimensions,
a result which confirms recent findings in Rayleigh-B\'enard (RB)
convection in porous media
\cite{hewitt2012ultimate, hewitt2014high}.
Surprisingly we find that, at given Rayleigh number, the Nusselt
number is larger in 2D than in 3D, while the opposite is observed
for RB convection \cite{hewitt2014high}.

\begin{figure}[h!]
\includegraphics[width=0.2\textwidth]{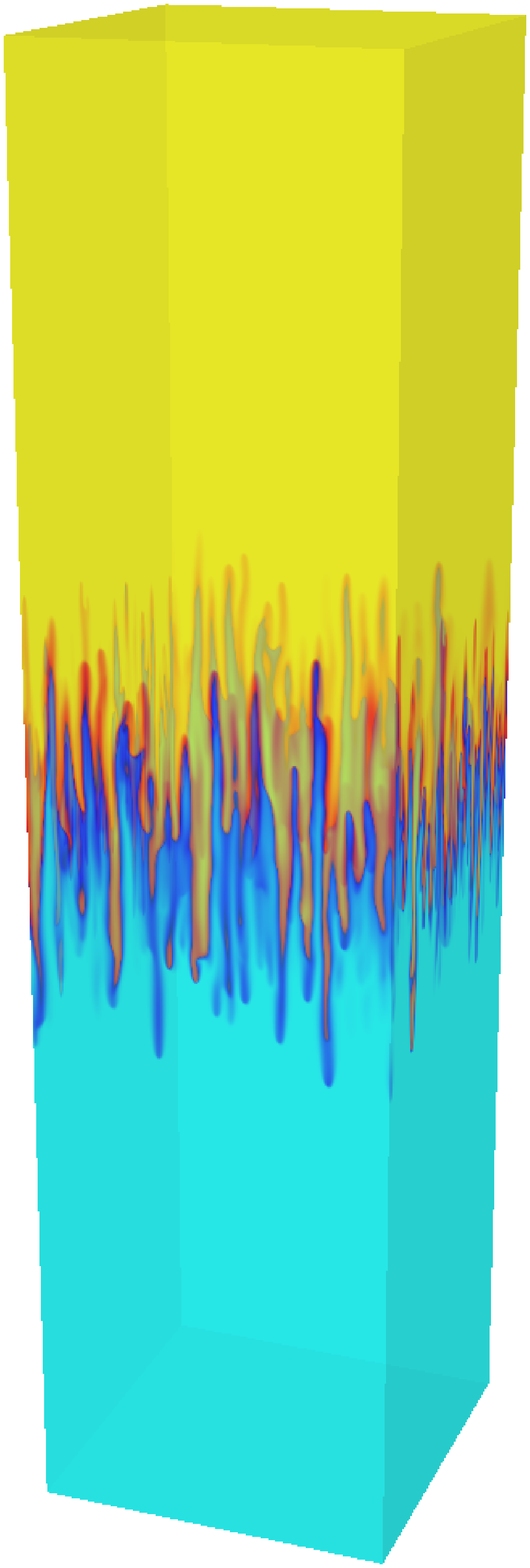}
\includegraphics[width=0.2\textwidth]{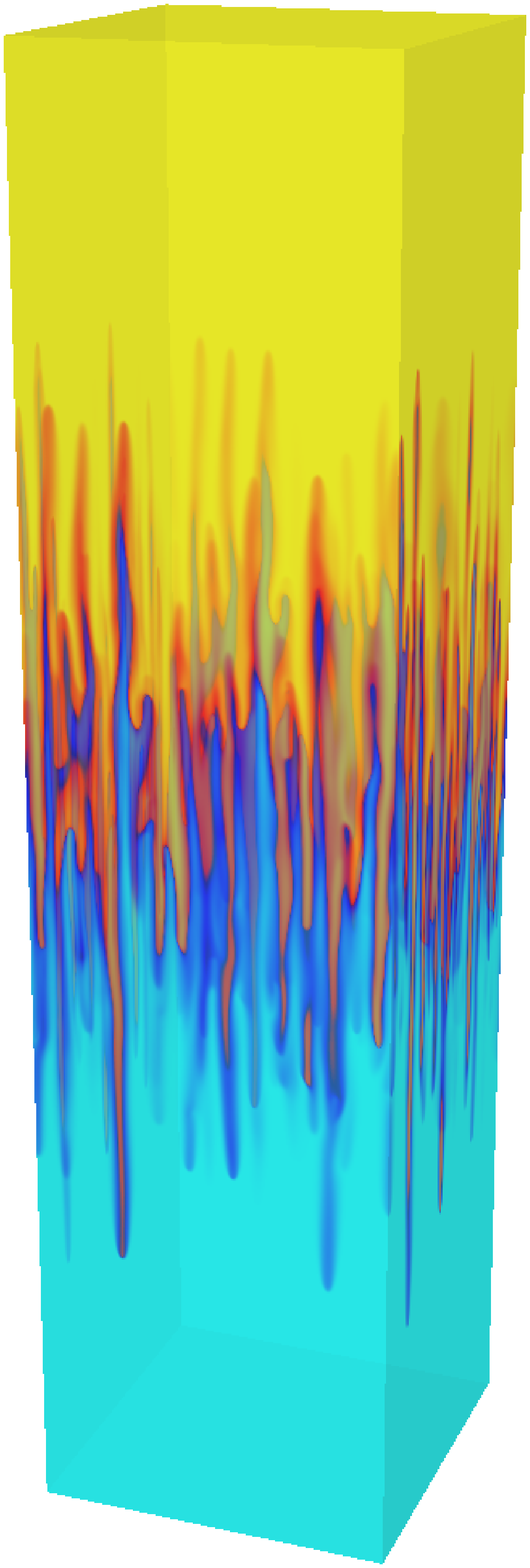} \\
\includegraphics[width=0.4\textwidth]{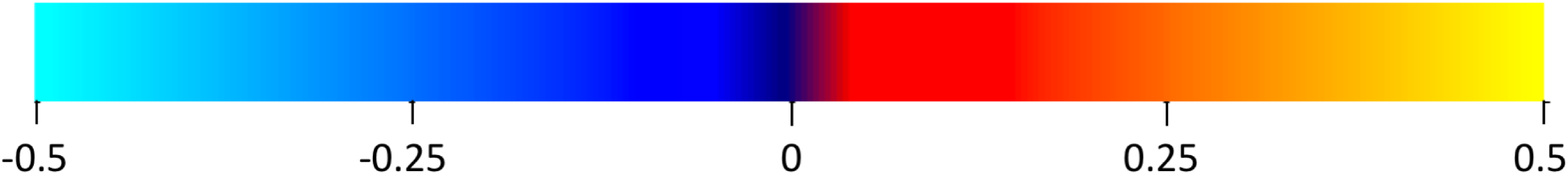}
\caption{Two snapshots of the density fields for the 3D simulation
at $Pe= 1.6 \times 10^4$  at times $t=2 \tau$ (left) and 
$t=4 \tau$ (right). Colors represent the relative density.
}
\label{fig1}
\end{figure}

We consider the Darcy's equation for a single-phase (miscible)
flow in an isotropic porous medium \cite{muskat1937flow}
in the Boussinesq approximation \cite{nield2006convection}
\begin{equation}
{\bf u} = {\kappa \over \mu \phi} (-{\bf \nabla} p + \rho {\bf g})
\label{eq:1}
\end{equation}
\begin{equation}
\partial_t \rho + {\bf u} \cdot {\bf \nabla} \rho = D \nabla^2 \rho
\label{eq:2}
\end{equation}
where ${\bf u}$ represents the velocity field (incompressible, i.e. 
${\bf \nabla} \cdot {\bf u}=0$) and $\rho$ is the density field.
In (\ref{eq:1}) $\kappa$ is the isotropic permeability, 
$\phi$ is the porosity,
$\mu$ is the viscosity (which is assumed constant), 
$D$ is the diffusivity coefficient 
and ${\bf g} = (0,0,-g)$ is the gravity acceleration. 
More generally, the flow ${\bf q}=\phi {\bf u}$ through 
a non-isotropic porous medium
can be expressed in terms of the the permeability tensor
$K_{ij}$ as $q_i = - K_{ij}(\partial_j p + \rho g \delta_{j3})/\mu$.
Here we consider for simplicity
the isotropic case $K_{ij} = \kappa \delta_{ij}$. 

The Rayleigh-Taylor setup is obtained by taking an initial unstable 
step density profile with $\rho=\rho_0 - \Delta \rho/2$ for 
$-H/2 \le z \le 0$ and $\rho=\rho_0 + \Delta \rho/2$ for 
$0 \le z \le H/2$, where  $\Delta \rho$ is the density jump.
$H$ and $L$ are the vertical and horizontal sizes of the box,
respectively. Given that the term $\rho_0 {\bf g}$ in Eq.~(\ref{eq:1}) 
can be absorbed in the hydrostatic pressure, 
in the numerical simulations we set $\rho_0 = 0$ without loss of
generality. The only parameter which controls the dynamics of the system
is the P\'eclet number $Pe = Lw_0/D$, where 
$w_0 = \frac{\kappa g}{\mu \phi} \Delta \rho$ 
is a characteristic velocity. 

We perform extensive numerical simulations of the
Darcy-Boussinesq equations (\ref{eq:1}) and (\ref{eq:2}) by means of 
pseudospectral code~\cite{boyd2001chebyshev} 
in both two and three dimensions.
The box sizes are $L=2 \pi$ and $H=4 L$ with a uniform
numerical grid at resolution $M \times M \times 4 M$.
The boundary conditions are periodic in all the directions
and we impose a zero-velocity mask around $z=\pm H/2$ (where the density
stratification is stable).
Time evolution is obtained by a second-order Runge-Kutta scheme
with explicit linear part and the code is fully parallelized in one
direction by using MPI libraries.
The time step is chosen to satisfy the Courant criterium.
In order to trigger the instability, a random perturbation
is added in a narrow region around the interface at $z=0$.
Being interested in investigating the possible asymptotic regime
and the effects of finite-Pe on the growth law of the mixing layer,
the values of $Pe$ considered in our study are the highest
permitted by the resolution of the simulation and are comparables
with values reported for $CO_2$ sequestration~\cite{backhaus2011convective}.

Three flows at different
diffusivities $D=4 \times 10^{-4}$, $D=2 \times 10^{-4}$ and
$D=10^{-4}$ (at resolutions $M=512$, $M=1024$ and $M=2048$ respectively)
are simulated, corresponding to P\'eclet numbers 
$Pe = 0.8 \times 10^{4}, Pe =1.6 \times 10^{4}$ and $Pe = 3.1 \times 10^{4}$.
In all cases the values of the diffusivity
is sufficiently large to resolve the initial instability
predicted by linear stability analysis, 
i.e. the maximum resolved wavenumber
$k_{max}=M/3$ is larger than $k^*={\sqrt{5}-2 \over 2 D} w_0$
\cite{wooding1962stability,trevelyan2011buoyancy} ($k_{max}/k^* = 1.16$).
During the evolution of the mixing layer, the most energetic mode move
to larger scales making this requirement less stringent 
\cite{celani2009phase,supmat}.
The results are averaged over an ensemble
of $2$ simulations for the 3D case
and $20$ simulations for the 2D case
with different initial random perturbations.
Numerical results are made dimensionless by using $L$ 
and $\tau=L/w_0$ as length and time units respectively.

\begin{figure}[h!]
\includegraphics[width=0.48\textwidth]{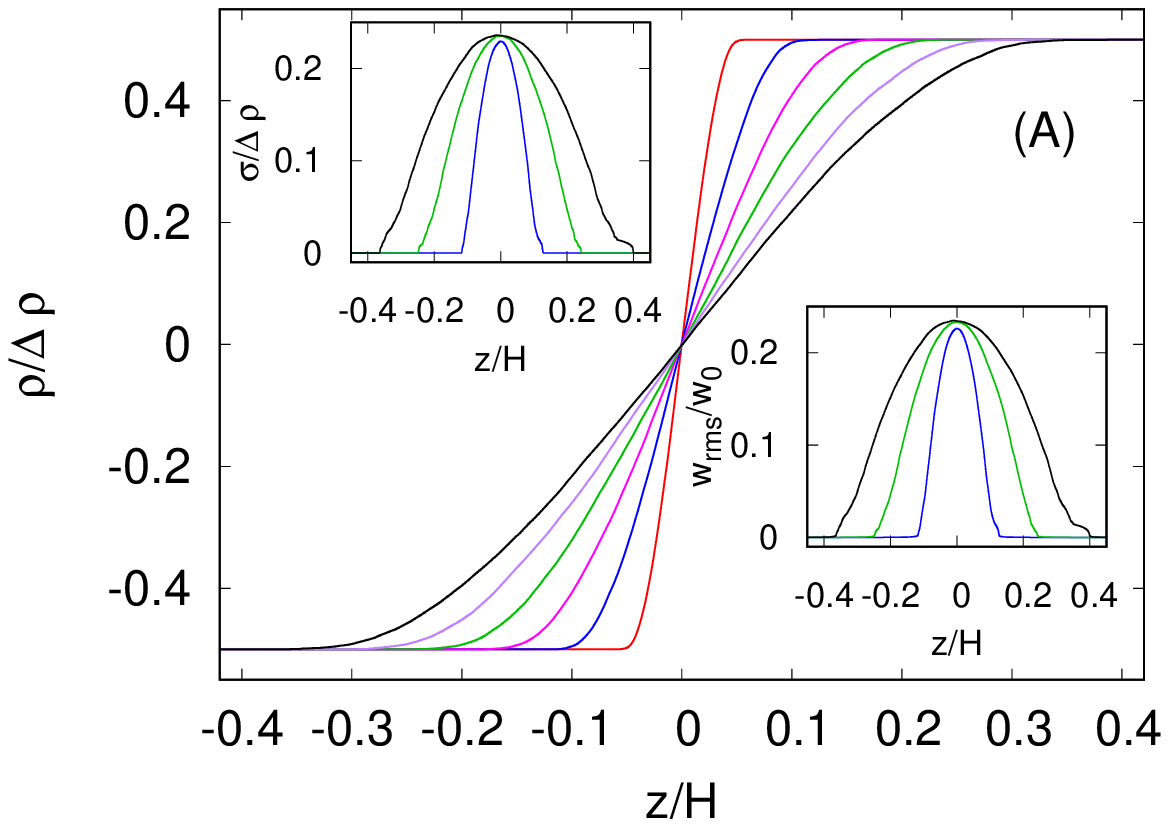}
\includegraphics[width=0.48\textwidth]{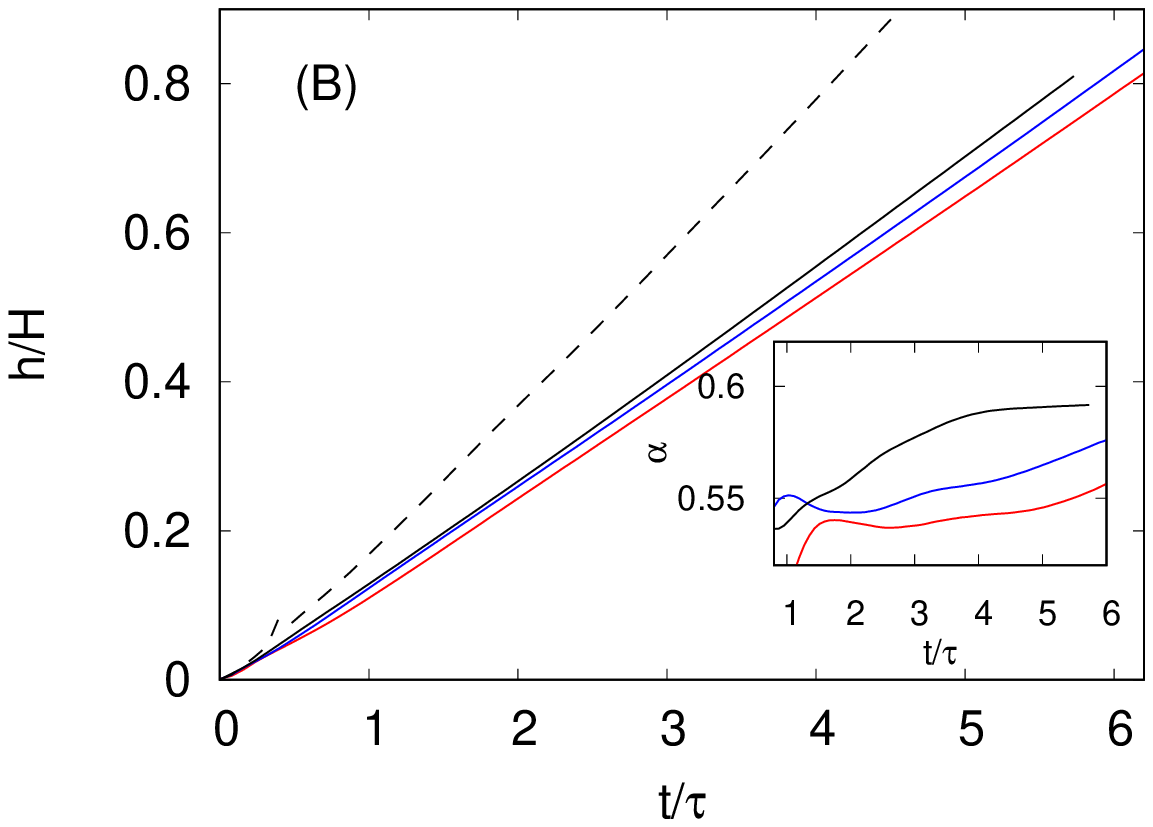}
\caption{
Left panel (A):
Mean density profiles $\overline{\rho}(z,t)$ at times 
$t=0.8 \tau$ (red), $t=1.6 \tau$ (blue), $t=2.4 \tau$ (magenta), 
$t=3.2 \tau$ (green), $t=4.0 \tau$ (violet) and $t=4.8 \tau$ (black).
Upper inset: density standard deviation 
$\sigma(z,t)=(\overline{\rho^2}-\overline{\rho}^2)^{1/2}$ at times 
$t=1.6 \tau$ (blue), $t=3.2 \tau$ (green) and $t=4.8 \tau$ (black).
Lower inset: rms vertical velocity profile $\overline{w}_{rms}(z,t)$ at times
$t=1.6 \tau$ (blue), $t=3.2 \tau$ (green) and $t=4.8 \tau$ (black).
Simulation in 3D at $Pe= 3.1 \times 10^4$.
Right panel (B): Time evolution of the width of the mixing layer $h(t)=2 z_1(t)$
for the simulations at $Pe= 0.8  \times 10^4$ (red), 
$Pe= 1.6  \times 10^4$ (blue) and $Pe= 3.1  \times 10^4$
(black). The black dashed line represents the two-dimensional case 
with $Pe= 3.1 \times 10^4$. 
Inset: time derivative $dh/dt$ normalized with $w_0$ to give the value
of the coefficient $\alpha = (dh/dt)/w_0 \simeq 0.59$.
}
\label{fig2}
\end{figure}

We first focus on the results of the 3D simulations.
In Fig.~\ref{fig1} we show two snapshots of the density field at 
different times in the evolution of the mixing layer. One observes that
heavy (light) plumes penetrating into light (heavy) domains 
become bigger and more elongated, suggesting the lack of self-similarity.
In spite of the complexity of the
mixing layer, the mean density vertical profile,
defined as the average over horizontal planes 
$\overline{\rho}(z,t)=\langle \rho({\bf x},t) \rangle_{xy}
= {1 \over L^2} \int \rho({\bf x},t) dx dy$,
displays a self-similar evolution with a 
close-to-linear gradient in the central part, as shown in Fig.~\ref{fig2}A. 
The upper inset in Fig.~\ref{fig2}A shows the density standard deviation 
$\sigma(z,t) \equiv \sqrt{\overline{\rho(z,t)^2} - \overline{\rho(z,t)}^2}$ inside
the mixing layer at three different times. While the 
region of non-zero $\sigma(z)$ extends in time following the mixing
layer, the peak value of $\sigma$ remains approximatively constant,
indicating a lack of complete homogenization.
We note that, in agreement with the Darcy's law (\ref{eq:1}), 
the profile of density standard deviation
is equal to the profile of the rms vertical velocity, i.e., 
$\overline{w}_{rms}(z,t) = w_0 \sigma(z,t)/\Delta\rho$ (see lower inset of 
Figure.~\ref{fig2}A). 

The observed simple density profile allows a precise determination 
of the size of the mixing layer (i.e. regions in which density is 
not uniform and velocity is non-zero) 
by using a non-linear diffusive model developed recently for turbulent 
RT mixing at high Reynolds numbers \cite{boffetta2010nonlinear}. 
The model is based on an eddy diffusivity which depends 
on the local density and predicts for the mixing layer the
polynomial profile
$\rho(z)={\Delta \rho \over 4} {z \over z_1} \left[3-
\left({z \over z1}\right)^2 \right]$ for $|z| \le z_1$, where $z_1$ 
represents half size of the mixing layer. 
We find that the above cubic expression fits very well the density profiles
shown in Fig.~\ref{fig2}A and allows to measure precisely the extension of the 
mixing layer as $h(t)=2 z_1(t)$.
We remark that other definitions of the width of the mixing layer 
have been used, based on local threshold value \cite{dalziel1999self}      
or global quantities \cite{cabot2006reynolds}. 
When applied to our data these methods give results similar to the
polynomial fit but with larger uncertainties.

The evolution of the mixing layer is shown in 
Fig.~\ref{fig2}B for the 3D simulations at different diffusivities. 
We notice that, at variance with the turbulent case where it follows an 
accelerated law \cite{boffetta2017incompressible}, here the mixing layer
grows linearly in time as $h(t)=\alpha w_0 t$ where
$\alpha$ is a dimensionless coefficient which measures the efficiency
of the process. 
The physical reason for the linear growth is that in porous media 
viscosity suppresses the inertial accelerating term and the velocity 
in the mixing layer reaches a constant value, which is proportional to
the dimensional estimate $w_0$. 
is constant in time (see lower inset of Fig.~\ref{fig2}A). 
From the time derivative of $h(t)$ which, for the simulations at highest
$Pe$, reaches a constant value at final times (see inset of
Fig.~\ref{fig2}B) we are able to measure the value of the 
dimensionless coefficient $\alpha = (dh/dt)/w_0 \simeq 0.59$ 
with still a possible weak dependence on $Pe$.

From the point of view of dimensional analysis,
the time derivative $\alpha = dh/dt$ and 
the peak value of the profile of $w_{rms}$ 
can be considered equivalent quantities. 
Nonetheless, the convergence to the asymptotic regime
for these two quantities may be different,
because $\alpha$ depends on the global evolution of the mixing layer,
while the peak of $w_{rms}$ is attained in a narrow central region.
In particular, our results indicate that the saturation of the 
peak value of $w_{rms}$ anticipates the constant-$\alpha$ regime.
Indeed, by comparing the insets of Figs.\ref{fig2}A and \ref{fig2}B,
we observe that in the simulation with the largest $Pe$
at time $t \sim 1.6 \tau$
the peak value of the profile of $w_{rms}$ has already become constant,
while $\alpha$ has not yet reached a constant plateau. 
Moreover, we have found that the peak of the profile of $w_{rms}$
reaches a constant value also in the 3D simulations at lower $Pe$ in
which the regime of constant $\alpha$ is not yet attained.   


The extension of the mixing layer defines the (time dependent)
Darcy-Rayleigh number of the
flow measuring the ratio of diffusive to buoyancy timescales and 
defined here as $Ra={w_0 h \over D}$. 
At large values of $Ra$ density
fluctuations are transported by buoyancy forces and the efficiency of
this process is quantified by the mass-transfer Nusselt number 
$Nu={\langle w \rho \rangle h \over D \Delta \rho}$ 
(brackets indicate average over the physical domain).
The classical argument for the dependence of $Nu$ over $Ra$ 
assumes that, at large values of $Ra$ the density flux become 
independent on $h$ and, for a porous medium, predicts an ``ultimate''
linear scaling $Nu \simeq Ra$.
The linear scaling is also a rigorous bound for Rayleigh-B\'enard
porous convection \cite{doering1998bounds,otero2004high} and
has been observed in numerical simulations of porous Rayleigh-B\'enard
convection both in 2D \cite{hewitt2012ultimate} and in 3D
\cite{hewitt2014high}.

\begin{figure}[h!]
\includegraphics[width=0.48\textwidth]{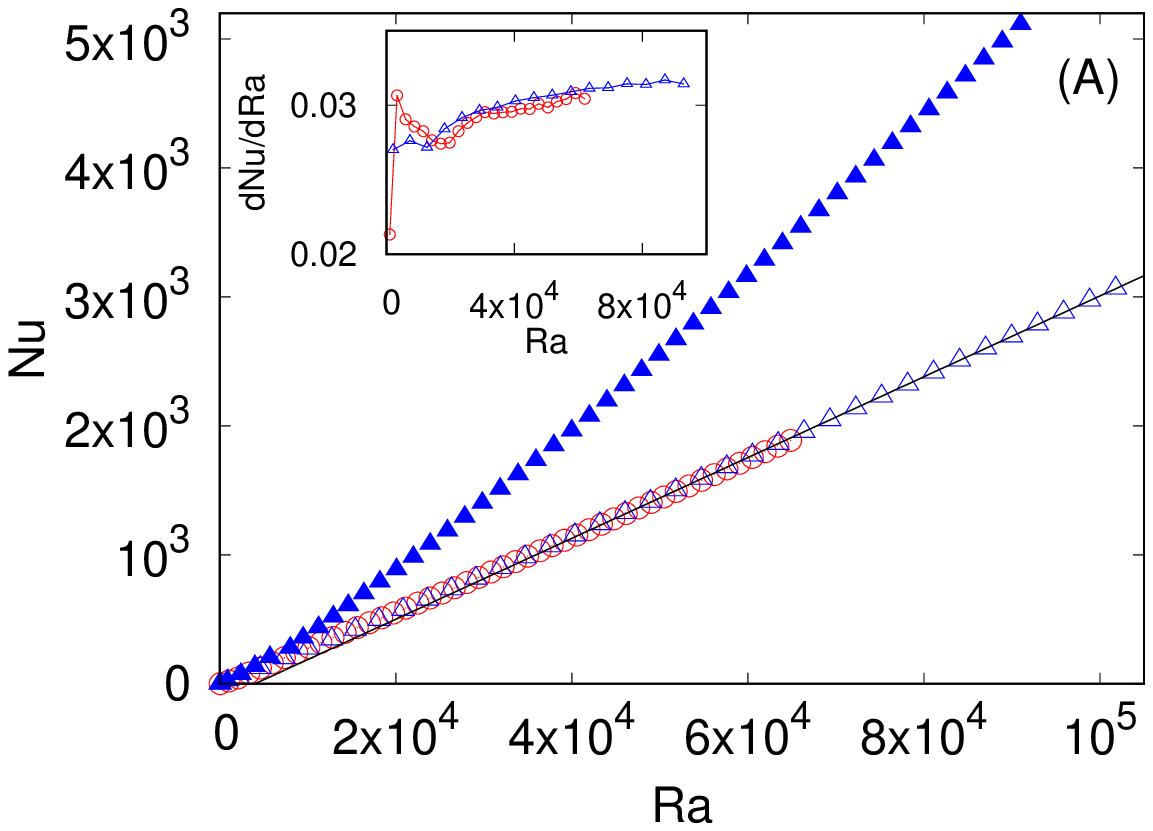}
\includegraphics[width=0.48\textwidth]{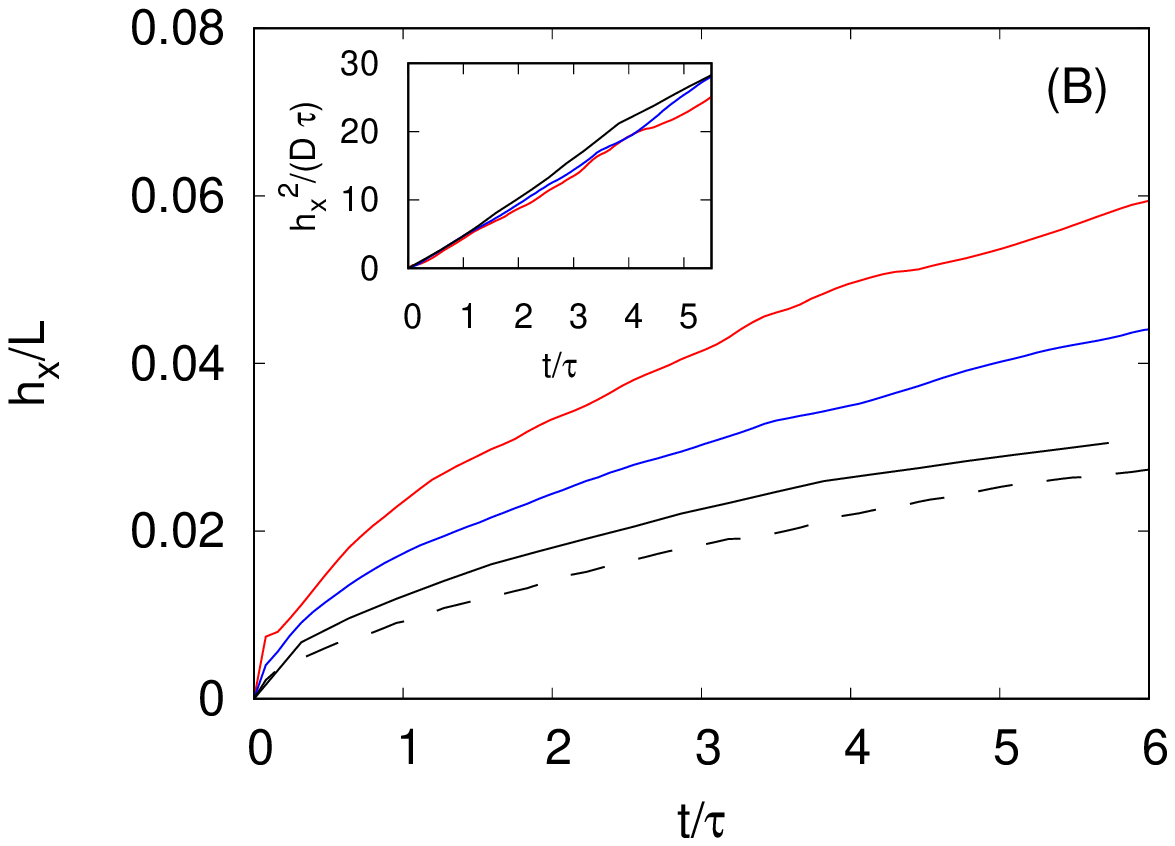}
\caption{
Left panel (A):
Nusselt number as a function of the Rayleigh number for the 
simulations at $Pe= 1.6 \times 10^4$ (red open circles) and 
$Pe= 3.1 \times 10^4$ (blue open triangles).
Blue filled triangles represent the results for the 2D case 
($Pe= 3.1 \times 10^4$).
The back line represents the linear dependence $Nu=0.031 Ra -124.9$.
Inset: derivative $d(Nu)/d(Ra)$ for the two 
three-dimensional cases (symbols as above).
Right panel (B): 
Time evolution of the horizontal correlation scale
$h_x(t)$ for the simulations at $Pe= 0.8  \times 10^4$ (red),
$Pe= 1.6 \times 10^4$ (blue) and $Pe= 3.1 \times 10^4$
(black).
The black dashed line represents the results for the two-dimensional
case with $Pe= 3.1 \times 10^4$.
Inset: square horizontal scale $h_x^2$ rescaled with $D \tau$
demonstrating diffusive growth.
}
\label{fig3}
\end{figure}

Figure~\ref{fig3}A shows the dependence of $Nu$ over $Ra$ for the 
3D runs at higher resolutions. For large values of $Ra$ a linear regime
is evident 
(and confirmed by the derivative shown in the inset)
for which $Nu \simeq 0.031 Ra$. This ultimate regime 
can be predicted for the case of RT mixing from the definition
of $Nu$ by assuming a constant 
$\langle w \rho \rangle = c w_0 \Delta \rho$ 
which gives $Nu = c Ra$. The coefficient $c \simeq 0.031$ 
measures the strength of the correlations between density fluctuations
and vertical velocity induced by the Darcy law. 


From Fig.~\ref{fig1} it is evident that density plumes become wider while 
elongating in the mixing layer. To quantify this process we compute 
the characteristic horizontal scale of plumes $h_x$ defined as the
first zero-crossing of the horizontal autocorrelation function 
of density $c(r,t)=\langle \rho(x+r,y,0,t) \rho(x,y,0,t) \rangle_{xy}/\langle
\rho(x,y,0,t)^2 \rangle_{xy}$, computed on the middle plane $z=0$. 
Figure~\ref{fig3}B shows the time evolution of $h_x(t)$ for the three 
different simulations with different $Pe$. 
We notice that here lines are in the opposite order with respect to 
Fig.~\ref{fig2}B, i.e. $h_x$ grows faster for the larger diffusivity. 
Indeed, as shown in the inset of Fig.~\ref{fig3}B, the 
evolution of horizontal scale is compatible with a diffusive growth 
$h_x(t)^2 \propto D t$. 
Nonetheless we remark that the diffusive growth of $h_x$ is the 
result of nonlinear processes including plume merging and coarsening 
and it is not a simple direct consequence of molecular diffusion.

The observation of a linear growth of the vertical scale (Fig.~\ref{fig2}B)
and the diffusive increase of the horizontal scale (Fig.~\ref{fig3}B)
indicates that two different processes are active in the dynamics of 
the mixing layer and that its evolution is not self-similar, at variance
with the turbulent case. 
The ballistic/diffusive evolution of the vertical/horizontal
scales has also been observed in quasi-2D Hele-Shaw cells
and in 2D simulations \cite{wooding1969growth,gopalakrishnan2017relative}
(with possible anomalous corrections for the vertical motion 
\cite{depaoli2019rayleigh}) and this suggests a possible universality
of porous RT mixing with respect to dimensionality. 
It is therefore interesting to directly compare our 3D results
with corresponding 2D cases.
To this aim, we have performed a set of numerical simulations for the 
two-dimensional version of (\ref{eq:1}) with the same resolutions and
diffusivities. The results in this case have been averaged over $10$ 
independent realizations. 

In Fig.~\ref{fig2}B we plot the evolution of the mixing layer for the 
two-dimensional simulations at the highest resolution and 
P\'eclet ($Pe= 3.1  \times 10^4$). 
After an initial regime in which the growth of $h(t)$ is faster than ballistic, 
at large times a linear growth is observed also in 2D but with
a coefficient $\alpha$ which is about $40\%$ larger than in the 
3D case ($\alpha_{2D} \simeq 0.84$).
The faster growth of the mixing layer in two dimensions is 
a consequence of density plumes which are more elongated 
and thinner than in the 3D case. 
This is confirmed by Fig.~\ref{fig3}B which shows that the horizontal 
correlation scale in the 2D case is smaller than in the 3D counterpart.

For the 2D simulations
at lower $Pe$ ($Pe= 0.8 \times 10^4 $ and $Pe= 1.6  \times 10^4$, 
not shown here) no linear regime is reached before $h$
reaches the size of the box and a faster-than-ballistic growth  
is observed, in agreement with recent studies \cite{depaoli2019rayleigh}.
Although the main goal of our work is not the precise
measurement of the scaling exponent $h(t) \sim t^\gamma$ in 2D,
we report that the value of $\gamma$ obtained in 2D at $Pe= 0.8 \times 10^4$
is close to $1.2$, in agreement with the value $\gamma = 1.208 \pm 0.008$
reported in \cite{depaoli2019rayleigh}.
Nevertheless, we have found that the value of the scaling exponent $\gamma$
reduces at increasing $Pe$, suggesting that it may be affected by finite $Pe$ 
and/or finite time corrections.  

The fact that the mixing layer in 2D grows faster than in 3D has not
{\it a priori} implications on the $Nu-Ra$ scaling since both quantities
depend on $h$. Nonetheless, as shown in Fig.~\ref{fig3}A,
dimensionality plays an important role also in the transfer of mass since
in 2D the Nusselt number at a given Rayleigh number is much larger 
than in 3D. For large values of $Ra$, we observe a linear $Nu-Ra$ 
relation also in 2D but with a coefficient which is about two
times larger than in the three-dimensional case.  
As discussed above, a larger value of the Nusselt number means
a larger correlation between the vertical velocity and the density 
fluctuations which, in the present case, is a consequence of 
more elongated and coherent plumes.
The scaling regime $Nu \simeq Re$ has been 
observed also in the case of Rayleigh-Benard (RB) convection in porous media
\cite{hewitt2012ultimate,hewitt2014high} but with an opposite dependence
of the numerical prefactor on the dimensionality:
in RB the coefficient in three dimensions is approximatively 
$40\%$ larger than in two dimensions\cite{hewitt2012ultimate,hewitt2014high},
while in RT we find that in two dimensions the
coefficient is almost two times larger than in 3D.

In conclusion, we have studied by means of extended direct numerical 
simulations of the Darcy-Boussinesq equation, the dynamics of 
Rayleigh-Taylor mixing in three dimensional porous media 
in a range of parameters relevant for applications such as $CO_2$ 
sequestration in saline aquifers \cite{backhaus2011convective}.
We have found that the growth of the mixing layer (i.e. plume elongation) 
follows accurately the dimensional linear prediction with a coefficient 
which weakly depends on the diffusivity. 
The horizontal width of density plumes is observed to grow much slower 
following a diffusive law. 
The mass-transfer Nusselt number is found to grow linearly with respect to 
the Rayleigh number, again in agreement with the dimensional prediction,
with a coefficient which is independent on the diffusivity (in the limit
of large Ra). 
The comparison of the results with the outcome of two-dimensional simulations
at the same diffusivity and resolution shows qualitative similarities but
important quantitative differences, therefore suggesting caution in
the use of 2D simulations and/or experiments for the modeling of 3D
applications 
such as, for example, the effectiveness $CO_2$ sequestration 
based on the reservoir parameters.
Moreover, our
results stimulate further studies aimed to investigate the transition 
between the 3D and the 2D regimes, which is expected to occur in
confined geometries at varying the relative extension of the horizontal scales.

\begin{acknowledgments}
We acknowledge support by the {\it Departments of Excellence} 
grant (MIUR) and the computing resources by HPC Center CINECA 
(IscrB-Pourun grant). We also thank A. Mazzino for useful
remarks.
\end{acknowledgments}

\bibliography{biblio}

\begin{thebibliography}{30}%
\makeatletter
\providecommand \@ifxundefined [1]{%
 \@ifx{#1\undefined}
}%
\providecommand \@ifnum [1]{%
 \ifnum #1\expandafter \@firstoftwo
 \else \expandafter \@secondoftwo
 \fi
}%
\providecommand \@ifx [1]{%
 \ifx #1\expandafter \@firstoftwo
 \else \expandafter \@secondoftwo
 \fi
}%
\providecommand \natexlab [1]{#1}%
\providecommand \enquote  [1]{``#1''}%
\providecommand \bibnamefont  [1]{#1}%
\providecommand \bibfnamefont [1]{#1}%
\providecommand \citenamefont [1]{#1}%
\providecommand \href@noop [0]{\@secondoftwo}%
\providecommand \href [0]{\begingroup \@sanitize@url \@href}%
\providecommand \@href[1]{\@@startlink{#1}\@@href}%
\providecommand \@@href[1]{\endgroup#1\@@endlink}%
\providecommand \@sanitize@url [0]{\catcode `\\12\catcode `\$12\catcode
  `\&12\catcode `\#12\catcode `\^12\catcode `\_12\catcode `\%12\relax}%
\providecommand \@@startlink[1]{}%
\providecommand \@@endlink[0]{}%
\providecommand \url  [0]{\begingroup\@sanitize@url \@url }%
\providecommand \@url [1]{\endgroup\@href {#1}{\urlprefix }}%
\providecommand \urlprefix  [0]{URL }%
\providecommand \Eprint [0]{\href }%
\providecommand \doibase [0]{http://dx.doi.org/}%
\providecommand \selectlanguage [0]{\@gobble}%
\providecommand \bibinfo  [0]{\@secondoftwo}%
\providecommand \bibfield  [0]{\@secondoftwo}%
\providecommand \translation [1]{[#1]}%
\providecommand \BibitemOpen [0]{}%
\providecommand \bibitemStop [0]{}%
\providecommand \bibitemNoStop [0]{.\EOS\space}%
\providecommand \EOS [0]{\spacefactor3000\relax}%
\providecommand \BibitemShut  [1]{\csname bibitem#1\endcsname}%
\let\auto@bib@innerbib\@empty
\bibitem [{\citenamefont {Boffetta}\ \emph {et~al.}(2009)\citenamefont
  {Boffetta}, \citenamefont {Mazzino}, \citenamefont {Musacchio},\ and\
  \citenamefont {Vozella}}]{boffetta2009kolmogorov}%
  \BibitemOpen
  \bibfield  {author} {\bibinfo {author} {\bibfnamefont {G}~\bibnamefont
  {Boffetta}}, \bibinfo {author} {\bibfnamefont {A}~\bibnamefont {Mazzino}},
  \bibinfo {author} {\bibfnamefont {S}~\bibnamefont {Musacchio}}, \ and\
  \bibinfo {author} {\bibfnamefont {L}~\bibnamefont {Vozella}},\ }\bibfield
  {title} {\enquote {\bibinfo {title} {{Kolmogorov scaling and intermittency in
  Rayleigh-Taylor turbulence}},}\ }\href@noop {} {\bibfield  {journal}
  {\bibinfo  {journal} {Phys. Rev. E}\ }\textbf {\bibinfo {volume} {79}},\
  \bibinfo {pages} {065301(R)} (\bibinfo {year} {2009})}\BibitemShut {NoStop}%
\bibitem [{\citenamefont {Chertkov}(2003)}]{chertkov2003phenomenology}%
  \BibitemOpen
  \bibfield  {author} {\bibinfo {author} {\bibfnamefont {Michael}\ \bibnamefont
  {Chertkov}},\ }\bibfield  {title} {\enquote {\bibinfo {title} {{Phenomenology
  of Rayleigh-Taylor turbulence}},}\ }\href@noop {} {\bibfield  {journal}
  {\bibinfo  {journal} {Phys. Rev. Lett.}\ }\textbf {\bibinfo {volume} {91}},\
  \bibinfo {pages} {115001} (\bibinfo {year} {2003})}\BibitemShut {NoStop}%
\bibitem [{\citenamefont {Celani}\ \emph {et~al.}(2006)\citenamefont {Celani},
  \citenamefont {Mazzino},\ and\ \citenamefont {Vozella}}]{celani2006rayleigh}%
  \BibitemOpen
  \bibfield  {author} {\bibinfo {author} {\bibfnamefont {A}~\bibnamefont
  {Celani}}, \bibinfo {author} {\bibfnamefont {A}~\bibnamefont {Mazzino}}, \
  and\ \bibinfo {author} {\bibfnamefont {L}~\bibnamefont {Vozella}},\
  }\bibfield  {title} {\enquote {\bibinfo {title} {{Rayleigh--Taylor Turbulence
  in Two Dimensions}},}\ }\href@noop {} {\bibfield  {journal} {\bibinfo
  {journal} {Phys. Rev. Lett.}\ }\textbf {\bibinfo {volume} {96}},\ \bibinfo
  {pages} {134504} (\bibinfo {year} {2006})}\BibitemShut {NoStop}%
\bibitem [{\citenamefont {Boffetta}\ \emph {et~al.}(2012)\citenamefont
  {Boffetta}, \citenamefont {De~Lillo},\ and\ \citenamefont
  {Musacchio}}]{boffetta2012bolgiano}%
  \BibitemOpen
  \bibfield  {author} {\bibinfo {author} {\bibfnamefont {G}~\bibnamefont
  {Boffetta}}, \bibinfo {author} {\bibfnamefont {F}~\bibnamefont {De~Lillo}}, \
  and\ \bibinfo {author} {\bibfnamefont {S}~\bibnamefont {Musacchio}},\
  }\bibfield  {title} {\enquote {\bibinfo {title} {{Bolgiano scale in confined
  Rayleigh--Taylor turbulence}},}\ }\href@noop {} {\bibfield  {journal}
  {\bibinfo  {journal} {J. Fluid Mech.}\ }\textbf {\bibinfo {volume} {690}},\
  \bibinfo {pages} {426} (\bibinfo {year} {2012})}\BibitemShut {NoStop}%
\bibitem [{\citenamefont {Sharp}(1984)}]{sharp1984overview}%
  \BibitemOpen
  \bibfield  {author} {\bibinfo {author} {\bibfnamefont {DH}~\bibnamefont
  {Sharp}},\ }\bibfield  {title} {\enquote {\bibinfo {title} {{An overview of
  Rayleigh-Taylor instability}},}\ }\href@noop {} {\bibfield  {journal}
  {\bibinfo  {journal} {Physica D}\ }\textbf {\bibinfo {volume} {12}},\
  \bibinfo {pages} {3--18} (\bibinfo {year} {1984})}\BibitemShut {NoStop}%
\bibitem [{\citenamefont {Boffetta}\ and\ \citenamefont
  {Mazzino}(2017)}]{boffetta2017incompressible}%
  \BibitemOpen
  \bibfield  {author} {\bibinfo {author} {\bibfnamefont {G}~\bibnamefont
  {Boffetta}}\ and\ \bibinfo {author} {\bibfnamefont {A}~\bibnamefont
  {Mazzino}},\ }\bibfield  {title} {\enquote {\bibinfo {title} {{Incompressible
  Rayleigh--Taylor turbulence}},}\ }\href@noop {} {\bibfield  {journal}
  {\bibinfo  {journal} {Annu. Rev. Fluid Mech.}\ }\textbf {\bibinfo {volume}
  {49}},\ \bibinfo {pages} {119} (\bibinfo {year} {2017})}\BibitemShut
  {NoStop}%
\bibitem [{\citenamefont {Zhou}(2017)}]{zhou2017rayleigh}%
  \BibitemOpen
  \bibfield  {author} {\bibinfo {author} {\bibfnamefont {Y}~\bibnamefont
  {Zhou}},\ }\bibfield  {title} {\enquote {\bibinfo {title} {{Rayleigh--Taylor
  and Richtmyer-Meshkov instability induced flow, turbulence, and mixing.
  I}},}\ }\href@noop {} {\bibfield  {journal} {\bibinfo  {journal} {Phys.
  Rep.}\ }\textbf {\bibinfo {volume} {720}},\ \bibinfo {pages} {1} (\bibinfo
  {year} {2017})}\BibitemShut {NoStop}%
\bibitem [{\citenamefont {Saffman}\ and\ \citenamefont
  {Taylor}(1958)}]{saffman1958penetration}%
  \BibitemOpen
  \bibfield  {author} {\bibinfo {author} {\bibfnamefont {P~G}\ \bibnamefont
  {Saffman}}\ and\ \bibinfo {author} {\bibfnamefont {G~I}\ \bibnamefont
  {Taylor}},\ }\bibfield  {title} {\enquote {\bibinfo {title} {{The penetration
  of a fluid into a porous medium or Hele-Shaw cell containing a more viscous
  liquid}},}\ }\href@noop {} {\bibfield  {journal} {\bibinfo  {journal} {Proc.
  R. Soc. Lond. A.}\ }\textbf {\bibinfo {volume} {245}},\ \bibinfo {pages}
  {312} (\bibinfo {year} {1958})}\BibitemShut {NoStop}%
\bibitem [{\citenamefont {Wooding}(1969)}]{wooding1969growth}%
  \BibitemOpen
  \bibfield  {author} {\bibinfo {author} {\bibfnamefont {R~A}\ \bibnamefont
  {Wooding}},\ }\bibfield  {title} {\enquote {\bibinfo {title} {{Growth of
  fingers at an unstable diffusing interface in a porous medium or Hele-Shaw
  cell}},}\ }\href@noop {} {\bibfield  {journal} {\bibinfo  {journal} {J. Fluid
  Mech.}\ }\textbf {\bibinfo {volume} {39}},\ \bibinfo {pages} {477} (\bibinfo
  {year} {1969})}\BibitemShut {NoStop}%
\bibitem [{\citenamefont {Huppert}\ and\ \citenamefont
  {Neufeld}(2014)}]{huppert2014fluid}%
  \BibitemOpen
  \bibfield  {author} {\bibinfo {author} {\bibfnamefont {Herbert~E}\
  \bibnamefont {Huppert}}\ and\ \bibinfo {author} {\bibfnamefont {Jerome~A}\
  \bibnamefont {Neufeld}},\ }\bibfield  {title} {\enquote {\bibinfo {title}
  {{The fluid mechanics of carbon dioxide sequestration}},}\ }\href@noop {}
  {\bibfield  {journal} {\bibinfo  {journal} {Annu. Rev. Fluid Mech.}\ }\textbf
  {\bibinfo {volume} {46}},\ \bibinfo {pages} {255} (\bibinfo {year}
  {2014})}\BibitemShut {NoStop}%
\bibitem [{\citenamefont {Ennis-King}\ and\ \citenamefont
  {Paterson}(2005)}]{ennis-king2005role}%
  \BibitemOpen
  \bibfield  {author} {\bibinfo {author} {\bibfnamefont {J}~\bibnamefont
  {Ennis-King}}\ and\ \bibinfo {author} {\bibfnamefont {L}~\bibnamefont
  {Paterson}},\ }\bibfield  {title} {\enquote {\bibinfo {title} {{Role of
  Convective Mixing in the Long-Term Storage of Carbon Dioxide in Deep Saline
  Formations}},}\ }\href@noop {} {\bibfield  {journal} {\bibinfo  {journal}
  {SPE Journal - SPE J}\ }\textbf {\bibinfo {volume} {10}},\ \bibinfo {pages}
  {349--356} (\bibinfo {year} {2005})}\BibitemShut {NoStop}%
\bibitem [{\citenamefont {Emami-Meybodi}\ \emph {et~al.}(2015)\citenamefont
  {Emami-Meybodi}, \citenamefont {Hassanzadeh}, \citenamefont {Green},\ and\
  \citenamefont {Ennis-King}}]{emami-meybodi2015convective}%
  \BibitemOpen
  \bibfield  {author} {\bibinfo {author} {\bibfnamefont {H}~\bibnamefont
  {Emami-Meybodi}}, \bibinfo {author} {\bibfnamefont {H}~\bibnamefont
  {Hassanzadeh}}, \bibinfo {author} {\bibfnamefont {C~P}\ \bibnamefont
  {Green}}, \ and\ \bibinfo {author} {\bibfnamefont {J}~\bibnamefont
  {Ennis-King}},\ }\bibfield  {title} {\enquote {\bibinfo {title} {{Convective
  dissolution of CO2 in saline aquifers: Progress in modeling and
  experiments}},}\ }\href@noop {} {\bibfield  {journal} {\bibinfo  {journal}
  {Int. J. Greenh. Gas Control}\ }\textbf {\bibinfo {volume} {40}},\ \bibinfo
  {pages} {238} (\bibinfo {year} {2015})}\BibitemShut {NoStop}%
\bibitem [{\citenamefont {Gopalakrishnan}\ \emph {et~al.}(2017)\citenamefont
  {Gopalakrishnan}, \citenamefont {Carballido-Landeira}, \citenamefont
  {De~Wit},\ and\ \citenamefont {Knaepen}}]{gopalakrishnan2017relative}%
  \BibitemOpen
  \bibfield  {author} {\bibinfo {author} {\bibfnamefont {Shyam~Sunder}\
  \bibnamefont {Gopalakrishnan}}, \bibinfo {author} {\bibfnamefont {Jorge}\
  \bibnamefont {Carballido-Landeira}}, \bibinfo {author} {\bibfnamefont {Anne}\
  \bibnamefont {De~Wit}}, \ and\ \bibinfo {author} {\bibfnamefont {Bernard}\
  \bibnamefont {Knaepen}},\ }\bibfield  {title} {\enquote {\bibinfo {title}
  {{Relative role of convective and diffusive mixing in the miscible
  Rayleigh-Taylor instability in porous media}},}\ }\href@noop {} {\bibfield
  {journal} {\bibinfo  {journal} {Phys. Rev. Fluids}\ }\textbf {\bibinfo
  {volume} {2}},\ \bibinfo {pages} {012501} (\bibinfo {year}
  {2017})}\BibitemShut {NoStop}%
\bibitem [{\citenamefont {De~Paoli}\ \emph {et~al.}(2019)\citenamefont
  {De~Paoli}, \citenamefont {Zonta},\ and\ \citenamefont
  {Soldati}}]{depaoli2019rayleigh}%
  \BibitemOpen
  \bibfield  {author} {\bibinfo {author} {\bibfnamefont {M}~\bibnamefont
  {De~Paoli}}, \bibinfo {author} {\bibfnamefont {F}~\bibnamefont {Zonta}}, \
  and\ \bibinfo {author} {\bibfnamefont {A}~\bibnamefont {Soldati}},\
  }\bibfield  {title} {\enquote {\bibinfo {title} {{Rayleigh-Taylor convective
  dissolution in confined porous media}},}\ }\href@noop {} {\bibfield
  {journal} {\bibinfo  {journal} {Phys. Rev. Fluids}\ }\textbf {\bibinfo
  {volume} {4}},\ \bibinfo {pages} {023502} (\bibinfo {year}
  {2019})}\BibitemShut {NoStop}%
\bibitem [{\citenamefont {Nakanishi}\ \emph {et~al.}(2016)\citenamefont
  {Nakanishi}, \citenamefont {Hyodo}, \citenamefont {Wang},\ and\ \citenamefont
  {Suekane}}]{nakanishi2016experimental}%
  \BibitemOpen
  \bibfield  {author} {\bibinfo {author} {\bibfnamefont {Y}~\bibnamefont
  {Nakanishi}}, \bibinfo {author} {\bibfnamefont {A.}~\bibnamefont {Hyodo}},
  \bibinfo {author} {\bibfnamefont {L.}~\bibnamefont {Wang}}, \ and\ \bibinfo
  {author} {\bibfnamefont {T}~\bibnamefont {Suekane}},\ }\bibfield  {title}
  {\enquote {\bibinfo {title} {{Experimental study of 3D Rayleigh–Taylor
  convection between miscible fluids in a porous medium}},}\ }\href@noop {}
  {\bibfield  {journal} {\bibinfo  {journal} {Adv. Water Res.}\ }\textbf
  {\bibinfo {volume} {97}},\ \bibinfo {pages} {224} (\bibinfo {year}
  {2016})}\BibitemShut {NoStop}%
\bibitem [{\citenamefont {Hewitt}\ \emph {et~al.}(2012)\citenamefont {Hewitt},
  \citenamefont {Neufeld},\ and\ \citenamefont {Lister}}]{hewitt2012ultimate}%
  \BibitemOpen
  \bibfield  {author} {\bibinfo {author} {\bibfnamefont {D~R}\ \bibnamefont
  {Hewitt}}, \bibinfo {author} {\bibfnamefont {J~A}\ \bibnamefont {Neufeld}}, \
  and\ \bibinfo {author} {\bibfnamefont {J~R}\ \bibnamefont {Lister}},\
  }\bibfield  {title} {\enquote {\bibinfo {title} {{Ultimate Regime of High
  Rayleigh Number Convection in a Porous Medium}},}\ }\href@noop {} {\bibfield
  {journal} {\bibinfo  {journal} {Phys. Rev. Lett.}\ }\textbf {\bibinfo
  {volume} {108}},\ \bibinfo {pages} {224503} (\bibinfo {year}
  {2012})}\BibitemShut {NoStop}%
\bibitem [{\citenamefont {Hewitt}\ \emph {et~al.}(2014)\citenamefont {Hewitt},
  \citenamefont {Neufeld},\ and\ \citenamefont {Lister}}]{hewitt2014high}%
  \BibitemOpen
  \bibfield  {author} {\bibinfo {author} {\bibfnamefont {D~R}\ \bibnamefont
  {Hewitt}}, \bibinfo {author} {\bibfnamefont {J~A}\ \bibnamefont {Neufeld}}, \
  and\ \bibinfo {author} {\bibfnamefont {J~R}\ \bibnamefont {Lister}},\
  }\bibfield  {title} {\enquote {\bibinfo {title} {{High Rayleigh number
  convection in a three-dimensional porous medium}},}\ }\href@noop {}
  {\bibfield  {journal} {\bibinfo  {journal} {J. Fluid Mech.}\ }\textbf
  {\bibinfo {volume} {748}},\ \bibinfo {pages} {879} (\bibinfo {year}
  {2014})}\BibitemShut {NoStop}%
\bibitem [{\citenamefont {Muskat}(1937)}]{muskat1937flow}%
  \BibitemOpen
  \bibfield  {author} {\bibinfo {author} {\bibfnamefont {M.}~\bibnamefont
  {Muskat}},\ }\href@noop {} {\emph {\bibinfo {title} {{The Flow of Homogeneous
  Fluids Through Porous Media}}}}\ (\bibinfo  {publisher} {McGraw-Hill},\
  \bibinfo {year} {1937})\BibitemShut {NoStop}%
\bibitem [{\citenamefont {Nield}\ and\ \citenamefont
  {Bejan}(2006)}]{nield2006convection}%
  \BibitemOpen
  \bibfield  {author} {\bibinfo {author} {\bibfnamefont {DA}~\bibnamefont
  {Nield}}\ and\ \bibinfo {author} {\bibfnamefont {A}~\bibnamefont {Bejan}},\
  }\href@noop {} {\emph {\bibinfo {title} {{Convection in Porous Media}}}}\
  (\bibinfo  {publisher} {Springer},\ \bibinfo {year} {2006})\BibitemShut
  {NoStop}%
\bibitem [{\citenamefont {Boyd}(2001)}]{boyd2001chebyshev}%
  \BibitemOpen
  \bibfield  {author} {\bibinfo {author} {\bibfnamefont {J~P}\ \bibnamefont
  {Boyd}},\ }\href@noop {} {\emph {\bibinfo {title} {{Chebyshev and Fourier
  Spectral Methods}}}}\ (\bibinfo  {publisher} {Dover Publication},\ \bibinfo
  {year} {2001})\BibitemShut {NoStop}%
\bibitem [{\citenamefont {Backhaus}\ \emph {et~al.}(2011)\citenamefont
  {Backhaus}, \citenamefont {Turitsyn},\ and\ \citenamefont
  {Ecke}}]{backhaus2011convective}%
  \BibitemOpen
  \bibfield  {author} {\bibinfo {author} {\bibfnamefont {Scott}\ \bibnamefont
  {Backhaus}}, \bibinfo {author} {\bibfnamefont {Konstantin}\ \bibnamefont
  {Turitsyn}}, \ and\ \bibinfo {author} {\bibfnamefont {R~E}\ \bibnamefont
  {Ecke}},\ }\bibfield  {title} {\enquote {\bibinfo {title} {{Convective
  Instability and Mass Transport of Diffusion Layers in a Hele-Shaw
  Geometry}},}\ }\href@noop {} {\bibfield  {journal} {\bibinfo  {journal}
  {Phys. Rev. Lett.}\ }\textbf {\bibinfo {volume} {106}},\ \bibinfo {pages}
  {104501} (\bibinfo {year} {2011})}\BibitemShut {NoStop}%
\bibitem [{\citenamefont {Wooding}(1962)}]{wooding1962stability}%
  \BibitemOpen
  \bibfield  {author} {\bibinfo {author} {\bibfnamefont {R~A}\ \bibnamefont
  {Wooding}},\ }\bibfield  {title} {\enquote {\bibinfo {title} {{The Stability
  of an Interface Between Miscible Fluids in a Porous Medium}},}\ }\href@noop
  {} {\bibfield  {journal} {\bibinfo  {journal} {J. Applied Math. Phys.
  (ZAMP)}\ }\textbf {\bibinfo {volume} {13}},\ \bibinfo {pages} {255} (\bibinfo
  {year} {1962})}\BibitemShut {NoStop}%
\bibitem [{\citenamefont {Trevelyan}\ \emph {et~al.}(2011)\citenamefont
  {Trevelyan}, \citenamefont {Almarcha},\ and\ \citenamefont
  {De~Wit}}]{trevelyan2011buoyancy}%
  \BibitemOpen
  \bibfield  {author} {\bibinfo {author} {\bibfnamefont {P~M~J}\ \bibnamefont
  {Trevelyan}}, \bibinfo {author} {\bibfnamefont {C}~\bibnamefont {Almarcha}},
  \ and\ \bibinfo {author} {\bibfnamefont {A}~\bibnamefont {De~Wit}},\
  }\bibfield  {title} {\enquote {\bibinfo {title} {{Buoyancy-driven
  instabilities of miscible two-layer stratifications in porous media and
  Hele-Shaw cells}},}\ }\href@noop {} {\bibfield  {journal} {\bibinfo
  {journal} {J. Fluid Mech.}\ }\textbf {\bibinfo {volume} {670}},\ \bibinfo
  {pages} {38} (\bibinfo {year} {2011})}\BibitemShut {NoStop}%
\bibitem [{\citenamefont {Celani}\ \emph {et~al.}(2009)\citenamefont {Celani},
  \citenamefont {Mazzino}, \citenamefont {Muratore-Ginanneschi},\ and\
  \citenamefont {Vozella}}]{celani2009phase}%
  \BibitemOpen
  \bibfield  {author} {\bibinfo {author} {\bibfnamefont {Antonio}\ \bibnamefont
  {Celani}}, \bibinfo {author} {\bibfnamefont {Andrea}\ \bibnamefont
  {Mazzino}}, \bibinfo {author} {\bibfnamefont {Paolo}\ \bibnamefont
  {Muratore-Ginanneschi}}, \ and\ \bibinfo {author} {\bibfnamefont {Lara}\
  \bibnamefont {Vozella}},\ }\bibfield  {title} {\enquote {\bibinfo {title}
  {{Phase-field model for the Rayleigh--Taylor instability of immiscible
  fluids}},}\ }\href@noop {} {\bibfield  {journal} {\bibinfo  {journal} {J.
  Fluid Mech.}\ }\textbf {\bibinfo {volume} {622}},\ \bibinfo {pages}
  {115--134} (\bibinfo {year} {2009})}\BibitemShut {NoStop}%
\bibitem [{sup()}]{supmat}%
  \BibitemOpen
  \href@noop {} {\ }\bibinfo {note} {See Supplemental Material at [...] for a
  linear stability analysis.}\BibitemShut {Stop}%
\bibitem [{\citenamefont {Boffetta}\ \emph {et~al.}(2010)\citenamefont
  {Boffetta}, \citenamefont {De~Lillo},\ and\ \citenamefont
  {Musacchio}}]{boffetta2010nonlinear}%
  \BibitemOpen
  \bibfield  {author} {\bibinfo {author} {\bibfnamefont {G}~\bibnamefont
  {Boffetta}}, \bibinfo {author} {\bibfnamefont {F}~\bibnamefont {De~Lillo}}, \
  and\ \bibinfo {author} {\bibfnamefont {S}~\bibnamefont {Musacchio}},\
  }\bibfield  {title} {\enquote {\bibinfo {title} {{Nonlinear diffusion model
  for Rayleigh-Taylor mixing}},}\ }\href@noop {} {\bibfield  {journal}
  {\bibinfo  {journal} {Phys. Rev. Lett.}\ }\textbf {\bibinfo {volume} {104}},\
  \bibinfo {pages} {034505} (\bibinfo {year} {2010})}\BibitemShut {NoStop}%
\bibitem [{\citenamefont {Dalziel}\ \emph {et~al.}(1999)\citenamefont
  {Dalziel}, \citenamefont {Linden},\ and\ \citenamefont
  {Youngs}}]{dalziel1999self}%
  \BibitemOpen
  \bibfield  {author} {\bibinfo {author} {\bibfnamefont {SB}~\bibnamefont
  {Dalziel}}, \bibinfo {author} {\bibfnamefont {PF}~\bibnamefont {Linden}}, \
  and\ \bibinfo {author} {\bibfnamefont {DL}~\bibnamefont {Youngs}},\
  }\bibfield  {title} {\enquote {\bibinfo {title} {{Self-similarity and
  internal structure of turbulence induced by Rayleigh--Taylor instability}},}\
  }\href@noop {} {\bibfield  {journal} {\bibinfo  {journal} {J. Fluid Mech.}\
  }\textbf {\bibinfo {volume} {399}},\ \bibinfo {pages} {1--48} (\bibinfo
  {year} {1999})}\BibitemShut {NoStop}%
\bibitem [{\citenamefont {Cabot}\ and\ \citenamefont
  {Cook}(2006)}]{cabot2006reynolds}%
  \BibitemOpen
  \bibfield  {author} {\bibinfo {author} {\bibfnamefont {WH}~\bibnamefont
  {Cabot}}\ and\ \bibinfo {author} {\bibfnamefont {AW}~\bibnamefont {Cook}},\
  }\bibfield  {title} {\enquote {\bibinfo {title} {{Reynolds number effects on
  Rayleigh--Taylor instability with possible implications for type Ia
  supernovae}},}\ }\href@noop {} {\bibfield  {journal} {\bibinfo  {journal}
  {Nat. Phys.}\ ,\ \bibinfo {pages} {562}} (\bibinfo {year}
  {2006})}\BibitemShut {NoStop}%
\bibitem [{\citenamefont {Doering}\ and\ \citenamefont
  {Costantin}(1998)}]{doering1998bounds}%
  \BibitemOpen
  \bibfield  {author} {\bibinfo {author} {\bibfnamefont {CR}~\bibnamefont
  {Doering}}\ and\ \bibinfo {author} {\bibfnamefont {P}~\bibnamefont
  {Costantin}},\ }\bibfield  {title} {\enquote {\bibinfo {title} {{Bounds for
  heat transport in a porous layer}},}\ }\href@noop {} {\bibfield  {journal}
  {\bibinfo  {journal} {J. Fluid Mech.}\ }\textbf {\bibinfo {volume} {376}},\
  \bibinfo {pages} {263--296} (\bibinfo {year} {1998})}\BibitemShut {NoStop}%
\bibitem [{\citenamefont {Otero}\ \emph {et~al.}(2004)\citenamefont {Otero},
  \citenamefont {Dontcheva}, \citenamefont {Johnston}, \citenamefont
  {Worthing}, \citenamefont {Kurganov}, \citenamefont {Petrova},\ and\
  \citenamefont {Doering}}]{otero2004high}%
  \BibitemOpen
  \bibfield  {author} {\bibinfo {author} {\bibfnamefont {J}~\bibnamefont
  {Otero}}, \bibinfo {author} {\bibfnamefont {LA}~\bibnamefont {Dontcheva}},
  \bibinfo {author} {\bibfnamefont {H}~\bibnamefont {Johnston}}, \bibinfo
  {author} {\bibfnamefont {RA}~\bibnamefont {Worthing}}, \bibinfo {author}
  {\bibfnamefont {A}~\bibnamefont {Kurganov}}, \bibinfo {author} {\bibfnamefont
  {G}~\bibnamefont {Petrova}}, \ and\ \bibinfo {author} {\bibfnamefont
  {CR}~\bibnamefont {Doering}},\ }\bibfield  {title} {\enquote {\bibinfo
  {title} {{High-Rayleigh-number convection in a fluid-saturated porous
  layer}},}\ }\href@noop {} {\bibfield  {journal} {\bibinfo  {journal} {J.
  Fluid Mech.}\ }\textbf {\bibinfo {volume} {500}},\ \bibinfo {pages}
  {263--281} (\bibinfo {year} {2004})}\BibitemShut {NoStop}%
\end{thebibliography}%

\end{document}